# THERMODYNAMIC RESOLUTION OF "GIBBS PARADOX"

Etkin V.A.

It is show that the Gibbs paradox is actually paralogism, viz. an erroneous statement sounding credible due to the statistic-mechanical interpretation of entropy as a measure of "any and all" irreversibility. As an alternative, the thermodynamic theory of irreversible mixing processes will be offered as allowing for the dependence of losses from the nature of gases being mixed.

Among the paradoxes of physics it can hardly be found one more equally as famous and enigmatic as the "Gibbs' paradox", viz. a statement that entropy builds up stepwise at changing from identical gases to a mixture of gases arbitrarily little distinguishable in their macro-physical and micro-physical properties (J. Gibbs, 1950). For a century this fact has not once become the object of investigation for both physicists and philosophers. To many of its investigators it seemed they could eventually explain why the entropy jumped with so queer independence from the degree and character of distingushability between the gases mixed, as well as why the notion "entropy of mixture" was inapplicable to identical gases. However, like the legendary sphinx that paradox has been thrashed over on pages of scientific books and magazines and has not yet left them till nowadays. As a result, the majority of its investigators have inclined to an opinion that the "Gibbs paradox is unsolvable on the plane of classic thermodynamics (B. Kedrov, 1969).

Other vistas open up to this problem from the positions of energodynamics. This chapter is dedicated to show that the Gibbs paradox is actually paralogism, viz. an erroneous statement sounding credible due to the statistic-mechanical interpretation of entropy as a measure of "any and all" irreversibility. As an alternative, the thermodynamic theory of irreversible mixing processes will be offered as allowing for the dependence of losses from the nature of gases being mixed.

# 1. Origin and Nature of «Gibbs' Paradox»

In his famous work "On Equilibrium of Heterogeneous Substances" (1875–1876) J. Gibbs set forth the following expression for the entropy of an ideal gas mixture:

$$S = \Sigma_k N_k(c_{vk}\ln T + R_k\ln \upsilon_k + s_{ok}), \qquad (1.1)$$

where $R_\mu$ – universal gas constant; $N_k$ – mole number of the $k^{th}$ substance; $c_{vk}$, $\upsilon_k$, $s_{ok}$ – isochoric heat capacity, partial volume and entropic constant of a mole of the $k^{th}$ substance, respectively [1].

Gibbs wrote this expression by analogy with Dalton law to which the pressure of an ideal gas mixture p is equal to the sum of partial pressures of the components $p_k$ ($p = \Sigma_k p_k$). When postulating this "similar principle regarding the gas mixture entropy", Gibbs made no mention whatever of what the individual characteristics $s_{ok}$ and $c_{vk}$ meant having them evidently assumed identical properties of a corresponding pure substance.

It is significant that Gibbs did not at all consider expression (1.1) as rigorously proven. He just assumed it would have been correct to initially accept this relationship as a fundamental equation describing an ideal gas mixture and then to substantiate the validity of such definition by properties which might have been derived from it. Applying this expression to the diffusion at the mixing of ideal gases as two separate masses, each of the gases initially occupying half a complete volume, he defines that the difference between the gas mixture entropy $S = (M_1 R_1\ln V + M_2 R_2\ln V)$ and the entropies before mixing constitutes the constant value

$$S - [M_1 R_1\ln(V/2) + M_2 R_2\ln(V/2)] = R_s \ln 2, \qquad (1.2)$$

where $R_s = M_1 R_1 + M_2 R_2$ – universal gas constant of the system as a whole.

Commenting on this result Gibbs notes, "It is significant that the value of this expression does not depend on kinds of the gases being mixed and degree of their difference"… since the "value $pV/T$

---

[1] Here, unlike the original, 1 mole is adopted as a quantity unit of the $k^{th}$ substance

is entirely defined by the number of molecules being mixed". Thus Gibbs himself traced nothing paradoxical in that result. However as investigators were studying the question, they encountered ever growing difficulties, which caused the "Gibbs' paradox" definition.

In voluminous literature dedicated to this question several standpoints are met regarding the nature of this paradox. A number of investigators (M. Leontovich, 1951; A. Samoilovich, 1955; P. Chambadal, 1963; S. Fraier, 1973, and others) identify the nature of the paradox with the impossibility of a limit change to identical gases in expression (1.1). In fact, (1.1) does not contain any parameters describing the difference between gases. Therefore it necessarily follows from this expression that entropy jumps when portions of the same gas are mixing. Gibbs himself having adhered to the Boltzmann's (probabilistic) interpretation of entropy saw nothing queer in that since a "mixture of the same-kind gas masses in principle differs from that of the different-kind gas masses" – for lack of information allowing, in principle at least, to separate them. However such an argument is evidently unacceptable from the positions of thermodynamics wherein the initial information of a system is restricted to definition of the thermal and caloric equations of state identical for ideal gases.

Some investigators refer to the Gibbs' theorem itself as a paradox. According to it, the entropy of a gas mixture is equal to the summary entropy of particular gases, each occupying the volume of the whole mixture at the same temperature. Gibbs substantiated this statement by an imaginary experiment on the reversible separation of gas mixtures thru semipermeable membranes. However an imaginary experiment may be used in thermodynamics to substantiate some statement providing its conclusions do not contradict theory only (K. Putilov, 1974). Therefore many investigators have not taken this "proof" as convincing. Herein multiple attempts are rooted to more rigorously prove the entropy additivity in the Gibbs' concept. The proofs of the said theorem insomuch offered are reduced to two main categories:

a) Method of semipermeable membranes, which, besides Gibbs himself, was used by Rauleich, 1875; L. Boltzmann ,1878; A. Wiedeburg, 1894; A. Bik,1903; B. Tamman, 1924; V. Nernst, 1929; P. Chambadal, 1963; B. Kedrov, 1969, and others.

b) Method of gas column in gravitational field, which, in particular, H. Lorenz (1927) and E. Schrödinger (1946) used.

All these methods were aimed to define the work of reversible mixture separation and eventually based, explicitly or implicitly, on the assumption of ideal membranes capable to provide the so-called "membrane equilibrium" (when the gas mixture pressure on one side of the membrane is counterbalanced by the partial pressure of one of the components on its other side). It is significant that in all imaginary experiments of such a kind after-investigators discovered a number of inaccuracies and disputable assumptions. Furthermore, from such reasoning based on "asymmetrical" semipermeable membranes (letting gas through in only one direction) results were obtained antipodal to the Gibbs' theorem (P. Chambadal, 1963). Those proved the additivity of component entropies found at the total pressure and temperature of the mixture when no entropy jump appeared at all.

A number of investigators (V. Luboshits, M. Podgoretsky, L. Gelfer, 1971, 1975; E. Gevorkian, R. Gevorkian, 1975, 1976) adopt a neutral attitude toward the Gibbs' paradox considering the entropy jump as quite natural for gases modifying their properties discretely. In this case it is unclear how far different (from a thermodynamic standpoint) such substances are as: isotopes (different in molecular mass, but equal in chemical properties), isobars (different in chemical properties, but equal in molecular mass), isomers (different only in their life span in excited state), optical antipodes (different optically due to different spatial grouping of molecules), etc.

Depending on investigators' attitude toward the Gibbs' paradox their interpretation of its "solution" varies. The overwhelming majority of investigators accept the statement of the mixing process entropy existing as a true one though sounding somewhat unusual and incredible. These investigators after Gibbs refer the entropy jump at the mixing to (1) principal impossibility of after-separation of the same-kind-gas mixture; (2) principle difference of physical and chemical properties of gases regarding the character of their variation (M. Planck, 1925); (3) discrete variation of atomic properties (E. Schrödinger, 1946; R. Kubo, 1970; A. Samoilovich, 1955; A. Sommerfeld, 1955; L. Terletsky, 1966; I. Bazarov, 1976); (3) density discontinuity at the mixing of various gases (P. Lelouchier, 1975); (4) some work to be done to create partial pressures (B. Kedrov, 1969), etc. Other investigators see the solution to the Gibbs' paradox in proving the fact the mixing process entropy depends on the degree the gases differ from each other (V. Luboshits, M. Podgoretsky, 1971; Y. Varshavsky, A. Sheinin, 1968; R. Gevorkian, E. Gevorkian, 1976), e.g., for a mixture of the same gases with a continuously equalizing composition.

It is just a minor part of investigators (J. Van der Vaals, F. Konstamm, 1911; P. Postma, 1927; P. Chambadal, 1963; A. Veinik, 1967; B. Casper, S. Fraier, 1973; M. Biot, 1977) including the author of this book (V. Etkin, 1973, 1991), who deny any entropy variations at the mixing of non-interacting gases, which is the most radical solution to the said paradox.

## 2. Thermodynamic Inadmissibility of the Gibbs' Paradox

There are a number of arguments evidencing that the Gibbs' paradox is actually paralogism, viz. an erroneous statement sounding credible in the conviction that entropy rises in any irreversible process. It is impossible to reproduce herein all arguments of various authors to substantiate this thesis. Therefore we will not go beyond those of the arguments which are of the methodological character and therefore sound most convincing.

Classic thermodynamics dealing with only closed systems is known to have been interested in only the variation of entropy, but not its magnitude. This entropy variation in course of some process does not depend on whether a system is considered as a mixture of the $k^{th}$ ideal gases or as a set of the same ideal gases separated with a movable heat-permeable membrane since from the thermodynamic standpoint all properties of a system are defined by exclusively its thermal and caloric state equations. This entropy variation for a system with an arbitrary and constant (in whole) composition is derived from a known expression:

$$\Delta S = \Sigma_k N_k R_\mu \ln T/T_o - \Sigma_k N_k R_\mu \ln p/p_o, \qquad (2.1)$$

where $T_o, p_o$ and $T, p$ – absolute temperature and pressure of the gas mixture at the beginning and at the end of whatever process, respectively.

It follows from the identity of equations (2.1) for a gas mixture or a set of pure gases that from the positions of thermostatics they have the same number of degrees of freedom. This number is defined by the number of independent variables of state and is equal to 2 for gases (thermal and mechanical degrees of freedom). Hence the thermodynamic properties of a system under consideration (either a gas mixture or a set of pure gases) are to the full extent characterized by two (thermal and caloric) state equations in the form:

$$pV = MRT; \quad U = C_v T, \qquad (2.2)$$

where $R$, $C_v = \Sigma_k N_k c_{vk}$ – universal gas constant and total isochoric heat capacity of the system, respectively, found experimentally without knowing its composition and studying properties of its components separately. At these conditions the definition of any other properties of the system, e.g. its composition, is superfluous. Gibbs himself admitted this fact having noted that for a constant-composition system the "state is completely characterized by the total mass $M$ so that the knowledge of the composition of a system is not the necessary condition to derive its state equations". Hence both the gas mixture entropy and the gas set entropy as functions of system state were defined by two parameters of system state ($T, p$ or $T, V$) and due to their constancy at isobaric-isothermal mixing remained unvaried. Thus from a thermostatic standpoint none process ran, the more so because none of energy effects were observed at that. In fact, the aggregate system with two degrees of freedom yet before mixing was in total (both thermal and mechanical) equilibrium. Gibbs quite realized that when noted that the "problems of thermodynamics refer just to the system states defined by such incomplete way". Therefore the Van der Vaals' standpoint (1911) is quite reasoned when he noted regarding the mixture of isotopes, "However, from a thermostatic standpoint the mixture of such substances should be considered as a single substance and, since entropy is defined as thermostatic, there are no reasons to talk of an entropy rise at diffusion".

Another contradiction is revealed in the thermodynamic approach to the problem of entropy constants' additivity at the mixing. In fact, Gibbs' equation (1.1) was based on the analogy of the fundamental equation with Dalton law

$$p = \Sigma_k M_k R_\mu T/V \qquad (2.3)$$

With regard to equation (1.1) Gibbs notes that this expresses a known principle, according to which the pressure of a gas mixture is equal to the sums of the pressures the components of this gas mixture would have providing they exist separately at the same volume and temperature. Thus Gibbs was explicitly based on additivity of entropy for each of the components as expressed by the relationship $S = \Sigma_k N_k s_k$.

Let us clarify now whether expression (1.1) complies with this relationship if the value $s_{ok}$ assumed constant in processes of system composition variations. The additivity (summability) of whatever extensive parameter is known to suppose its specific value does not depend on mass. In other words, additive values are homogeneous mass functions, i.e. comply with Euler's theorem which in entropy application has the form:

$$\partial S_k(N_k)/\partial N_k = S_k(N_k)/N_k . \qquad (2.4)$$

Expression (2.4) is compatible with (1.1) when the derivative $(\partial S_k/\partial N_k)$ does not depend on $N_k$. However, it is easy to see that at $s_{ok} =$ const the entropy $S_k$ of a particular component does not meet this requirement. It really follows from (1.1):

$$\partial S_k(N_k)/\partial N_k = s_k(N_k) - R_\mu, \qquad (2.5)$$

i.e. is a function of $N_k$. Hence the Gibbs' assumption that $s_{ok} =$ constant and the same before and after mixing is groundless.

One more contradiction revealed by B. Kedrov (1969) is that the variation of ideal gas mixture entropy depends on the process path. Let a vessel A contain a mixture of two moles $H_2$ (hydrogen) and two moles $Cl_2$ (chlorine). The vessel is kept in darkness so that a chemical reaction within it is practically "inhibited". Let us separate the mixture with a partition into two equal parts and initiate in one of them by exposure to light an isobaric-isothermal chemical reaction resulting in generating two HCl molecules. As a result of the chemical reaction, the entropy of this part of the system will vary by some value $\Delta S_x$. Let us now remove the partition between the first half of a system containing the $H_2+Cl_2$ mixture and the second half containing the HCl gas. Owing to the fact that the gases are different in both halves of the vessel, according to the Gibbs' mixing theory the system entropy will rise by some value $\Delta S_{mix}$. Now let us expose the both halves of the vessel to light – the reaction will develop further with one more HCl mole generated and the entropy further varied by $\Delta S_x$. The total entropy variation in the three said processes resulted in two HCl moles generated is equal to $2\Delta S_x + \Delta S_{mix}$. However, the same two HCl moles could be obtained by exposing the mixture to light as a whole, i.e. without its separation followed by mixing. In this case the gas mixture entropy would evidently vary by only a value of $2\Delta S_x$. Since the initial and final state of the system and the heat effect of the reaction are the same in both variants, a contradiction is present.

Further contradiction is revealed when using the mixing process entropy to calculate the exergy (capability for technical work) losses of the systems workability at the mixing of gases. According to the Gibbs' mixing theory which does not involve any parameters characterizing differences between gases, the workability loss – $\Delta E_{mix}$ at the mixing of substances featuring ideal gas state (i.e. complying with the Clapeyron equation) is defined by exclusively the mixing process entropy $\Delta S_{mix}$ and the environmental temperature $T_0$ and does not depend on the chemical nature of the substances being mixed:

$$-\Delta E_{mix} = T_o \Delta S_\text{см}. \qquad (2.6)$$

Let us consider, however, a fuel cell, to which electrodes, e.g., oxygen and hydrogen are fed separately under a minor pressure (so that their state would not differ from ideal gas). The chemical affinity in reversible fuel cells is known to be realized in the form of electric current work which is theoretically equal to the chemical affinity value. Now let us mix oxygen and hydrogen partly or completely before feeding them to the fuel cell electrodes, i.e. let us feed not pure gases, but some oxy-hydrogen mixture. The similar experiments have repeatedly been conducted and are known to have led to drop of the voltage developed by the fuel cell down to total disappearance of current in the external circuit. Hence the actual loss of fuel cell capability for work depends on the nature of gases being mixed (their chemical affinity) and reaches the 100% value when the reaction becomes thermodynamically irreversible. This example is even more remarkable because allows to distinguish the losses at the mixing and chemical transformation. From this example it follows that the major losses arise not during mixing of gases, but in subsequent chemical reaction which due to this becomes thermodynamically irreversible. Therefore it would be more correct to refer in this case not to losses at the mixing, but rather losses due to mixing.

Inapplicability of the Gibbs-obtained result shows up when as well estimating the capability for work of whatever substance which concentration differs from its environmental concentration[1]. Classic thermodynamics according to the Gibbs' theory gives the following equation for the exergy of working medium in flow $Ex_p$ (Szargut J., Petela R.,1968):

$$Ex_p = H - H_o - T_o(S - S_o) + R_c T_o \Sigma_k \ln p_k/p_{ok}, \qquad (2.7)$$

where $H, H_o$ и $S, S_o$ – enthalpy and entropy of a working medium (gas) at initial state and at equilibrium with the environment, respectively; $p_k$, $p_{ok}$ – partial pressures of the $i^{th}$ substance in initial mixture and in the environment, respectively.

The last term of this expression defines the so-called "chemical" (more exactly, concentration) exergy caused by the difference between partial pressures of the $k^{th}$ substances in the system $p_k$ and the environment $p_{ok}$. This may supposedly be realized with the help of semipermeable membranes which allow isothermally expanding the gas from the pressure $p_k$ to $p_{ok}$ in an expanding machine with heat obtained from the environment and useful external work done.

As follows from this expression, the unit mass exergy for an ideal gas does not depend on its chemical nature and tends to infinity as its environmental concentration is decreasing. Inapplicability of such a conclusion is evident.

---

[1] This allows in-principle constructing an engine using this concentration difference

Thus estimating the Gibbs' approach to the gas mixing theory it has to be admitted that this gives no answer to not only the most important question about criteria of difference or identity of gases being mixed, but either about the theoretical value of work to be done to separate the mixture. The experiment shows that the less the difference between the mix components, the more the work on gas separation, even to say nothing of the test hardware imperfection. In particular, when producing nuclear fuel by separating a gas mixture containing 99,3% $U^{238}F_6$ and 0,7% $U^{235}F_6$, it is theoretically required (allowing for mixing process entropy) 0.023kWh of energy per 1kg of the second component. However, the actual energy consumption amounts to $1.2 \cdot 10^6$ kWh, i.e. approximately fifty million times as much (J. Ackeret, 1959). Thus the Gibbs' mixing process entropy can not serve as a basis for even approximate estimation of the theoretical mix separation work.

### 3. Entropy Reference Point Shift in Mixing Process as Entropy "Jump" Reason

Far from all investigators of the Gibbs' paradox have related this with the change to investigation of open system in the Gibbs' concept. In fact, considering the ideal gas mixture entropy or ideal gas set entropy from the positions of "pre-Gibbs" thermodynamics of closed systems as a composition function, i.e. $S = S(T,p,N_k)$, the entropy at the mixing will remain unvaried since the temperature, pressure and mole numbers $N_k$ (masses $M_k$) of all system components remain unvaried at that. In other words, for the diffusion process in its intrinsic meaning as the concentration equalization in a closed system the entropy remains unvaried despite the irreversibility of this process.

Let us consider now the isobaric-isothermal mixing process in the Gibbs' concept as a composition variation in each of the open subsystems due to the exchange of the $k^{th}$ substances among them (i.e. a mass transfer among them). In this case, because additional degrees of freedom appear (related to the $k^{th}$ substances exchange), the exact differential of the system entropy becomes:

$$dS = (\partial S/\partial T)dT + (\partial S/\partial p)dp + \Sigma_k (\partial S/\partial N_k)dN_k . \qquad (3.1)$$

The first and the second partial derivatives in this expression are defined at constant composition and mass of the whole system ($M_k$ = const, $\Sigma_k M_k$ = const) and may be found from the joint equation of the 1st and the 2nd laws of thermodynamics for closed systems. In particular, for the derivative of entropy with respect to temperature from (1.1) via caloric state equation (2.2) the following forms may be found:

$$(\partial S/\partial T)_{V,N} = C_v/T \: ; \: (\partial S/\partial P)_{T,N} = -MR_c/P. \qquad (3.2)$$

As for the derivative $(\partial S/\partial N_k)_{T,V}$, this can not be defined only based on the laws of thermodynamics for closed systems. To define this, relationship (1.1) should be applied, which gives:

$$(\partial S/\partial N_k)_{T,V} = s_k, \qquad (3.3)$$

where $s_k$ – partial molar entropy of the $k^{th}$ component, i.e. the value characterizing the entropy $S$ rise in an open system when one mole of the $k^{th}$ substance enters in it at isobaric-isothermal conditions, whereas the mole number of other, $j^{th}$ substances, does not vary ($k \neq j$).

Thus for open systems the exact differential of the entropy $S = S(T,p,N_k)$ is:

$$dS = (C_v/T)dT + (MR/V)dV + \Sigma_k \: s_k dN_k. \qquad (3.4)$$

Integrating this expression from an initial arbitrary state with the entropy $S_0$ assuming $C_v$ constant and allowing for the relationship $dV/V = d\upsilon/\upsilon$ evident at $N$ = const gives:

$$S = C_v \ln T + NR_\mu \ln \upsilon + \Sigma_k \int s_k dN_k + S_o \: . \qquad (3.5)$$

This expression differs from (1.1) offered by Gibbs by the third term appeared on the right-hand side and describing the entropy variation at gas mixing. This is what Gibbs neglected when having integrated equation (3.4) with respect to only the variables $T$ and $V$. As a result, he defined the entropy to an accuracy of some function of state $S_o(N_k)$, i.e. obtained not the entropy constant, but some function of composition as the sum of the two last terms on the right-hand side of expression (12.3.5). This value necessarily varies at the mixing in course of diffusion of the $k^{th}$ gases.

It may be easily shown that at the mixing to the Gibbs' concept both the mix entropy and its reference point, i.e. the $S_o(N_k)$ value, vary simultaneously and equally.

Let us consider the same gas set system which Kedrov used in his imaginary experiment when mixing $H_2$ and $Cl_2$. Let us assume after Gibbs that $s_{ok} = 0$, $S_o = 0$. Let us further transfer the system via an arbitrary quasi-static (e.g. isochoric) process to a state with a temperature of $T_1$ and volume $V_1$. The entropy of the system will then rise by a value $\Delta S_{0-1} = \Sigma_k N_k R_\mu \ln T_1/T_o$. Now let us remove the partition and provide the isobaric-isothermal mixing where the system volume remains unvaried ($V_1=V_2$), whereas the entropy according to the Gibbs' theory rises by some value $\Delta S_{mix}$ and becomes equal to $S_2 = S_1 + \Delta S_{mix}$. Then let us cool the mixture obtained down to a state with a temperature of $T_3 = T_0$. The entropy will subsequently decrease by a value of $\Delta S_{2-3} = -\Delta S_{0-1}$. The system has again returned to the state with the same temperature and volume, however, now the entropy in this state (which we adopt as initial) is equal to $S_3 = S_2 + \Delta S_{2-3} = S_o + \Delta S_{mix}$. Thus the gas mixture entropy value at the reference point parameters has varied by the exact mixing process entropy value! In other words, *in a diffusion process as the subject of the Gibbs' concept not the entropy itself experiences a jump, but its reference point!* Other result could hardly be expected since the Gibbs' mixing process entropy does not depend on temperature and hence is the same both in its current state and an arbitrary reference point. As a matter of fact, applying expression (1.1) to an arbitrary reference point for entropy any unbiased investigator would arrive at a conclusion that this reference point has as well experienced the same jump. Thus the entropy jump, should it really take place, equally relates to also the entropy reference point since this jump depends on only the ratio of mixed gas volumes before and after mixing. However, to justify the fallacy, it should be noted that in the days of Gibbs the problem of the entropy reference point selection and the entropy magnitude definition, which has eventually led to the third law of thermodynamics, did not yet exist. This is the circumstance that, in our view, engendered the Gibbs' paradox. It should seem so that the after-investigators could not have omitted the fact. They knew the third law of thermodynamics. It is this law that defines the reference point for entropy of all condensed substances. As a matter of fact, according to the third law of thermodynamics "as temperature is approaching the absolute zero, the entropy of any equilibrium system in isothermal processes ceases depending on whatever thermodynamic state parameters and to the limit $T = 0$ adopts a constant value, the same for all systems, which may be assumed as zero" (I. Bazarov, 1976). Therefore such a shift of the entropy reference point comes into antagonism with the third law of thermodynamics which reads that the entropy of whatever equilibrium system at the absolute zero temperature adopts a constant value, the same for all systems, which may be assumed as zero. Thus the entropy reference point jump as ensuing from the mixing process to the Gibbs' concept leads to a conflict with the third law of thermodynamics. This fact, which is, as far we know, beyond other investigators' comments, exactly reveals the paralogism of the Gibbs' paradox.

From the above-mentioned it becomes clear why some of the investigators, based on imaginary experiments, came to the necessity of calculating the mixture entropy from the mixture total volume, whereas the others – from the total pressure and temperature of the mixture. The fact is that both of these standpoints are equally valid and applicable since neither of them leads to a jump of entropy at the mixing of non-interacting gases.

Hence from the positions of energodynamics as well the Gibb's conclusion of a step rise of entropy at the mixing of ideal gases appears as an erroneous statement caused by the arbitrariness in choosing the reference point of entropy for open systems, i.e. by the violation of the third law of thermodynamics. This result shows that in thermodynamics the Gibbs' paradox does not take place whatever meaning is read into it. As for statistic and informational entropies, the jump is not something paradoxical here since the number of possible permutation of particles depends on whether the gases are considered identical or distinguishable.

## 4. Thermodynamic Theory of Mixing Processes

The gas mixing theory must give a solution to two problems, viz. the definition of the useful (free) energy of a particular mixture and the work needed for its separation. The answer the Gibbs' theory gives solves neither of them. This is a challenge to approach the issue from the positions of energodynamics. Energodynamics considers the mixing as an irreversible process of equalizing the concentrations of the components all over the system volume while keeping their number unvaried for the system as a whole. This corresponds to the strict import of the word "diffusion" (from the Latin "diffusio" meaning spreading). Such a redistribution of components involves the variation of the moment of the $k^{th}$ substances' distribution $\mathbf{Z}_k$. As any irreversible process, this may be maintained from outside, e.g. by feeding the $k^{th}$ substances across the system borders. However, this is independent of the transfer of these substances across the system borders (diffusion across the borders), i.e. of the diffusion to the Gibbs' concept. The latter, for the avoidance of mishmash, we have named above the *selective mass transfer* of a system. The considered process involves a variation of the mass of the whole system $M = \Sigma_k M_k$ or the mole number in it $N = \Sigma_k N_k$. Unlike this, the diffusion leaves $M$ and $N$ unvaried. In fact, the mixture component distribution may be equalized also in the absence of a substance transfer across the system borders as it occurs, e.g. under the influence of electric or centrifugal fields. Thus the mixing is a specific qualitatively distinguishable irreversible process irreducible to other processes. This can not be approached by the "adaptation" to whatever other process.

The diffusion coordinate for the $k^{th}$ independent substance is the moment of its distribution $\mathbf{Z}_k$, while the motive force $\mathbf{X}_k$ – the component potential concentration gradient which is defined by the uniqueness conditions of the mixing process and, above means the negative gradient of diffusive, osmotic, etc. potential. The diffusion process is irreversible and involves thermal and bulk effects resulting in internal sources of entropy and volume appeared. To find these sources, let us represent the entropy S and the mixture volume V in terms of their partial molar values $s_k$ and $\upsilon_k$, respectively:

$$S = \Sigma_k N_k s_k;\ V = \Sigma_k N_k \upsilon_k,\qquad(4.1)$$

and obtain:

$$\Delta S_{mix} = \Sigma_k N_k (s_k - s_{ko}); \quad \Delta V_{mix} = \Sigma_k N_k (\upsilon_k - \upsilon_{ko}). \tag{4.2}$$

The value $(s_k - s_{ko})$ in this equation characterizes the variation of entropy of the $k^{th}$ pure substance $s_{ko}$ in the mixing process, while the associated heat

$$q_k^* = T(s_k - s_{ko}) \tag{4.3}$$

– thermal effects arising at the mixing of interacting components[1]. These effects are caused by the available partial molar entropy (defined by an actual increment in mixture entropy $s_k$ at isobaric-isothermal input of the unit $k^{th}$ substance) and by the entropy $s_{ko}$ superinduced from outside by one mole of the pure $k^{th}$ substance. This means that the above value belongs to the thermodynamic function of mixing. Similarly the difference $\Sigma_k N_k(\upsilon_k - \upsilon_{ko})$ characterizes the bulk effects arising at the mixing of interacting components, while the value

$$w_k^* = p(\upsilon_k - \upsilon_{ko}) \tag{4.4}$$

– cubic strain work involved in these effects. For non-interacting substances the thermal and the bulk effects $(s_k - s_{ko}; \upsilon_k - \upsilon_{ko})$ at input of the $k^{th}$ substance are absent ($q_k^* = w_k^* = 0$), which confirms the above conclusion that the ideal gas mixing process entropy is absent.

Thus the thermal and the bulk effects $q_k^*$ and $w_k^*$ described by expressions (4.3) and (4.4) may serve as a measure of mixing process irreversibility. These relationships allow answering all questions raised above in the thermodynamic mixing theory. According to them the theoretical work on mixture separation depends on the nature of gases being separated and for non-interacting gases becomes zero along with the difference $s_k - s_{ko}$. This circumstance

---

[1] In the theory of irreversible processes (TIP) the value $q_k^*$ is introduced as one of the transport factors $L_{kj}$, while its interpretation as the energy transported by one mole of the $k^{th}$ substance in the absence of temperature gradient becomes possible just "a posteriori" (by experimental results)

leads to a necessity to distinguish the vector and the scalar stages of an energy dissipation process. Let us elucidate this by example of a fuel cell realizing the chemical affinity of reagents in the form of electric work. If the reagents, before fed to the fuel cell electrodes, are allowed to be completely mixed, the fuel cell emf is known to fall down to zero. The mixing process has the vector nature according to the tensor order of its coordinates $\mathbf{Z}_k$. Though this stage is irreversible and involves thermal and bulk mixing effect, it just a little changes the value of reagents' chemical affinity. However, this stage results in that the subsequent homogeneous chemical reaction becomes thermodynamically irreversible with the charges not separated and useful work done, but with a heat released in amount equivalent to this work. Thus the capability for work is lost as a result of the spatial homogeneity spontaneously set in for the chemically reacting system, but not the subsequent chemical reaction itself. The chemical energy appeared to have already been dissipated by the beginning of this reaction and there was nothing for it but to pass into the random energy (heat). In other words, the chemical energy of the reacting mixture appeared to have already been less ordered than the initial energy of the spatially separated reagents. It is referred to the fact that instead of a macro-heterogeneous system we have got the micro-heterogeneous one differing in just the structure (configuration) of the molecules and atoms comprising it. However, this is not yet the heat energy of reaction products! Thus we come to a necessity to distinguish between the macro-physical stage of a dissipation process involving the disappearance of system spatial heterogeneity (it is expressed mathematically by "scalarization" of the process, i.e. loosing its vector nature) and its microphysical stage associated with scaling heterogeneity down to a mere random form of energy obtained. Using equation of displacement vectors' balance the first stage may feature the value

$$W^d = \Sigma_i \mathbf{F}_i \cdot d_s \mathbf{r}_j, \tag{4.5}$$

which may be named the "dissipation micro-work".

The second, micro-physical, stage of the dissipation process reflects the destruction of a chemically reacting system. This stage as well involves the internal "disggregation work" (R. Clausius) done. However, this work features already the scalar character and must be described as the "dissipation micro-work". This is the work that defines the decline of free energy of a chemically reacting mixture. In a general case this category of work should include not only the scalar chemical reaction work, but as well any work associated with further "disordering" of the substance (rearrangement of its molecular, crystal, cluster, etc. structure).

Hence the standard affinity of a homogeneous chemical reaction is partly consumed to prepare the reaction mixture (vector stage of the process not obeying the stoichiometric proportions), and then – to run the reaction itself (V. Etkin, 1991). This may be the reason why the standard affinity of a reaction can not play the part of its thermodynamic force (G. Gladyshev, 1991). Anyway, the possibility to distinguish between the thermal effects of mixing and subsequent chemical reaction may serve as the incentive to further investigations.